\documentclass[pdflatex,sn-aps]{sn-jnl}
\catcode`\|=12\relax

 \usepackage{tikz}
\usepgfmodule{decorations} 
\usetikzlibrary{calc}

\usepackage{siunitx}
\usepackage{graphicx}
\usepackage[retainorgcmds]{IEEEtrantools}
\newcommand{\lambdabar}{{\mkern0.75mu\mathchar '26\mkern -9.75mu\lambda}}

\title{On the experimental description of neutron resonances}

\author[1]{\fnm{Julien} \sur{Gibelin}}\email{gibelin@unicaen.fr}

\affil*[1]{ \orgname{Universit\'e de Caen Normandie, ENSICAEN, CNRS/IN2P3, LPC Caen UMR6534}, \postcode{F-14000} \city{Caen}, \country{France}}

\keywords{nuclear structure, resonance properties, neutron-rich}

\begin{document}

\abstract{We present a collection of simple derivations for the
  neutron-induced resonance cross-sections. These formulae are
  commonly used to experimentally describe the fundamental properties
  of resonances for neutron-rich nuclei far from stability and to
  describe unbound nuclei. The main goal of this article is to
  illustrate their dependencies with basic observables in order to
  discuss the pertinence of experimental approaches in the derivation
  of their properties, especially for ``$N$-body'' resonances.  }

\maketitle

\section{Introduction}

The advent of new generation radioactive beam facilities has provided
the ability of experimentally measuring resonances far from the valley
of stability, and in particular on the neutron-rich side. It also
allows neutron unbound nuclei to be produced and studied, and their
fundamental characteristics to be deduced. Physicists widely use a set
of simple formulae to extract their properties. However theses
equations have intrinsic limitations than can lead to misunderstanding
when they are compared with their theoretical description. We present
in the following sections a simple approach for their derivation,
keeping their expression dependent on only simple tabulated functions.
We then illustrate and discuss their behavior with experimentally
accessible observables and parameters.

\section{Derivation of the main equations}

\subsection{Scattering cross-section}

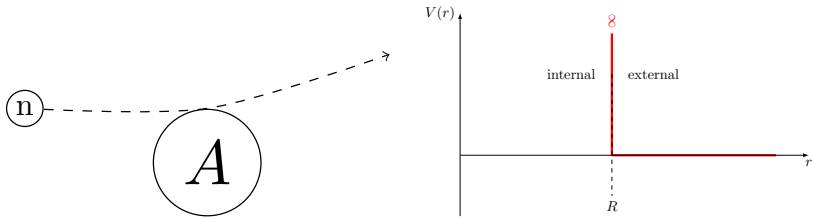
\begin{figure}[!ht]
  \centering
  \resizebox{0.45\textwidth}{!}{\begin{tikzpicture}
\draw (0,0) circle (.59) node[scale=2] {$A$};
\draw[dashed,->]  plot [smooth] coordinates {(-2,0.6) (0,0.6) (2.,1.2)};
\draw[black,fill=white] (-2,0.6) circle (.2) node {n} ;
\end{tikzpicture} }
  \resizebox{0.45\textwidth}{!}{\begin{tikzpicture}
\draw[ultra thick,red] (3.7,0)--+(0,3) node[rotate=90,right] {$\infty$};
\draw[ultra thick,red] (3.7,0)--+(4,0) ;
\draw[dashed] (3.7,2)--+(0,-3) node[below] {$R$};
\draw (3.7,2)  node[left] {internal\phantom{O}} node[right] {\phantom{O}external};
\draw[-latex]  (0,-1.5)--+(0,5) node[left] {$V(r)$};
\draw[-latex] (0,0)--+(8.5,0) node[below] {$r$};
\end{tikzpicture}
  }
   \caption{Schematic (left) and potential (right) of the reaction
     considered here.}
  \label{fig:th:neut:frag}
\end{figure}

As we will see later, describing neutron states in the continuum and
in particular unbound (neutron-rich) nuclei, comes down at first order
to describing the resonant scattering of a neutron on a nucleus
(figure \ref{fig:th:neut:frag}).

If \textit{a priori} we should consider a dynamical evolution of the
wave packet describing the diffusion of the neutron on the nuclear
potential of the nucleus of interest, we can show that the size of the
corresponding wave packet is large enough to be decomposed into
standing waves \cite{Austern}.

A large part of the formalisms and derivations presented below come
from the books of Blatt and Weisskopf \cite{blatt} or Satchler
\cite{satchler:intro} as well as from the $R$--matrix formalism
\cite{rmatrix,rmatrix:simple}.

We can represent the incident beam (here the neutron) by a
plane wave:
\begin{equation}
  \exp(i \vec k \vec r ) = \exp( i k z)
\end{equation}
the wave number being calculated from the energy $E$ of the system
fragment plus neutron:
\begin{equation}
  \vec k = \frac{\mu \vec v}{\hbar } \Rightarrow k = \frac{\sqrt{2\mu
     E}}{\hbar}
  \qquad \mathrm{with} \qquad \mu = \frac{M_{frag} M_n}{M_{frag} + M_n}
  \label{eq:def:k}
\end{equation}
We develop this plane wave on the basis of spherical harmonics
$\mathcal{Y}_{\ell,m}(\theta)$, taking into account that we have a
cylindrical symmetric problem and that only the harmonics with $m=0$
come into play:
\begin{IEEEeqnarray}{rClCl}
  \exp(i k z ) & = & \sum\limits_{\ell=0}^{\infty} \mathcal{A}_\ell(r) \mathcal{Y}_{\ell,0}(\theta)\\
  \text{with } \mathcal A_l(r) & = & \int \mathcal{Y}_{\ell,0}^\ast(\theta) \exp \left(i k r
    \cos \theta\right) d\Omega & = & i^\ell
  \sqrt{4\pi(2\ell+1)} j_\ell(kr)\nonumber
\end{IEEEeqnarray}

and $j_p$ the spherical Bessel function of first kind and order $p$.
The expression of the latter simplifies at large distances (here
$k r \gg \ell$): $j_p (x) \to \frac{\sin(x-\frac{p\pi}{2})}{x}$ and by
recalling that $\sin(x) = \frac{e^{ix} - e^{-ix}}{2i}$, we get:
\begin{equation}
     \exp(i k z ) \approx \frac {\sqrt \pi }{kr}
     \sum\limits_{\ell=0}^{\infty} \sqrt{2\ell+1}\, i^{\ell+1}
     \Big\{\underbrace{\exp{\left[-i\ \left( kr- \frac{\pi}{2} \ell
            \right) \right] }}_\text{incoming wave}
     \underbrace{- \exp{\left[+i\ \left( kr- \frac{\pi}{2} \ell \right)
         \right] }}_\text{outgoing wave}\Big
     \}\mathcal{Y}_{\ell,0}
\end{equation}

However, this expression describes an unperturbed wave. As we are
trying to describe the scattering of this wave on the nuclear
potential of the fragment, we have to modify this expression for the
outgoing wave, by allowing the latter to be scaled by a complex factor
$\eta_\ell$:
\begin{equation}
  \label{eq:psi:waveplane}
  \psi \left( \vec r\right) = \frac {\sqrt \pi }{kr}
  \sum\limits^{\infty}_{\ell=0} \sqrt{2\ell+1}\, i^{\ell+1}
  \Big\{\underbrace{\exp{\left[-i\left( kr- \frac{\pi}{2} \ell
        \right) \right] }}_\text{incoming wave}
  \underbrace{- \eta_\ell\exp{\left[+i\left( kr- \frac{\pi}{2} \ell \right)
      \right] }}_\text{outgoing wave}\Big\}\mathcal{Y}_{\ell,0}
\end{equation}
The scattered wave is then obtained by
$\psi_\text{sc} = \psi - \exp(i k z ) $, whose integration of the flux
gives the scattering cross-section:
\begin{equation}
  \sigma_{sc,\ell}=\pi\lambdabar^{2}\left(2\ell+1\right)\left\|1-\eta_\ell\right\|^{2}
\label{eq:xs:scatterin}
\end{equation}
 The integration of the flux of the total wave function $\psi$ 
allows us to deduce the expression of the reaction cross-section:
\begin{equation}
  \sigma_{r,\ell}=\pi \lambdabar^{2}\left( 2\ell+1\right) \left( 1-\left\| \eta_\ell \right\|^{2}\right)
\label{eq:xs:reaction}
\end{equation}
A  cross-section being necessarily positive, the expression of
$\sigma_{r,\ell}$ implies $\|\eta_\ell\|\leq 1$.

\subsection{Expression of energy shift and penetrability}

Let us now address the reaction of interest.  The radial Schrödinger
equation gives :
\begin{equation}
  \frac {d^2u_\ell} {dr^2}+\left[ k^2-\frac {\ell\left( \ell+1\right) }
    {r^2}-\frac {2 \mu} {\hbar^2}V(r)\right] u_\ell = 0
\end{equation}
with $\mu$ the reduced mass of the system and $V(r)$ the Coulomb
potential. Within the hard sphere approximation, and since neutrons
are uncharged, $V(r) =0$ for $r>R$. We can thus write outside the
nucleus:
\begin{equation}
  \frac {d^2u_\ell} {dr^2}+\left[ k^2-\frac {\ell\left( \ell+1\right) }
    {r^2}\right] u_\ell = 0
\end{equation}
With the so-called \emph{regular} (which goes to $0$ when $r$ goes to
$0$) and \emph{irregular} solutions, respectively noted $F_\ell(r)$
and $G_\ell(r)$\footnote{Riccati-Bessel functions, solutions of the
  differential equation
  \( x^2 \frac{d^2 y}{dx^2} + \left[x^2 - n (n+1)\right] y = 0\). To
  get back to the present problem simply put $n=\ell, x=kr$ and
  $y=u_\ell\left(kr\right)$.}, written:
\begin{IEEEeqnarray}{rClCl}
  F_\ell( r) & = & \phantom{-} \sqrt{\frac{\pi kr}{2}} J_{\ell+1/2}\left(k r\right)\\
  G_\ell( r) & = & - \sqrt{\frac{\pi kr}{2}} Y_{\ell+1/2}\left(k
                 r\right) & = & \left(- 1\right)^\ell \sqrt{\frac{\pi
                   kr}{2}}J_{-\left(\ell+1/2\right)}\left(kr\right)
\end{IEEEeqnarray}

where $J_p$ and $Y_p$ are the Bessel functions of order $p$,
respectively of first and second kind.

We then have:
\begin{IEEEeqnarray}{lrClCl}
  \text{-- incoming term: }\quad& u_\ell^{(+)} &=& G_\ell(r) + i
  F_\ell(r) &=& \sqrt{\frac {\pi kr} 2} H^{{(2)}}_{\ell+1/2}(kr)\\
  \text{-- outgoing term: }\quad& u_\ell^{(-)} &=&
  \overline{u}_\ell^{(+)} &=& \sqrt{\frac {\pi kr} 2} H^{{(1)}}_{\ell+1/2}(kr)
\end{IEEEeqnarray}

with $H_{p}^{(1,2)}(z)$ the Hankel functions of order $p$, of the first and
second kind. Outside the nucleus, the radial part of the wave function
is a linear combination of the incoming and outgoing terms:
\begin{equation}
  u_\ell = \alpha\, u_\ell^{(+)} + \beta\, u_\ell^{(-)}
\end{equation}

As at large distances the wave function must converge to the
shape found in \eqref{eq:psi:waveplane}:

\begin{equation}
  \lim_{r \to \infty} u_\ell = \frac {\sqrt \pi }{kr}  \sqrt{2\ell+1}\, i^{\ell+1}
  \Big\{\exp{\left[-i\left( kr- \frac{\pi}{2} \ell \right) \right] }
  - \eta_\ell\exp{\left[+i\left( kr- \frac{\pi}{2} \ell \right)
    \right] }\Big\}
\end{equation}

By recalling that the asymptotic forms of the Hankel functions are:
\begin{displaymath}
  H_p^{1}(z) \to \sqrt{\frac{2}{\pi
     z}}\exp\left[+i\left(z-\frac{p\pi}{2}-\frac{\pi}{4}\right)\right]
  \text{ and } H_p^{2}(z) \to \sqrt{\frac{2}{\pi
     z}}\exp\left[-i\left(z-\frac{p\pi}{2}-\frac{\pi}{4}\right)\right]
\end{displaymath}

this implies:
\begin{IEEEeqnarray}{rCl}
 \lim_{r \to \infty} u_\ell^{(+)} &=& +i\ \exp\left[-i\left(kr-\frac{\pi}{2}\ell\right)\right]\\
 \lim_{r \to \infty} u_\ell^{(-)} &=& -i\ \exp\left[+i\left(kr-\frac{\pi}{2}\ell\right)\right]
\end{IEEEeqnarray}

and thus we have:
\begin{IEEEeqnarray}{rCl}
  \alpha & = & i^{}\sqrt{ 2+1}\frac {\sqrt {\pi
    }} {kr} \\
  \beta & = & -\eta_\ell \ \alpha
\end{IEEEeqnarray}
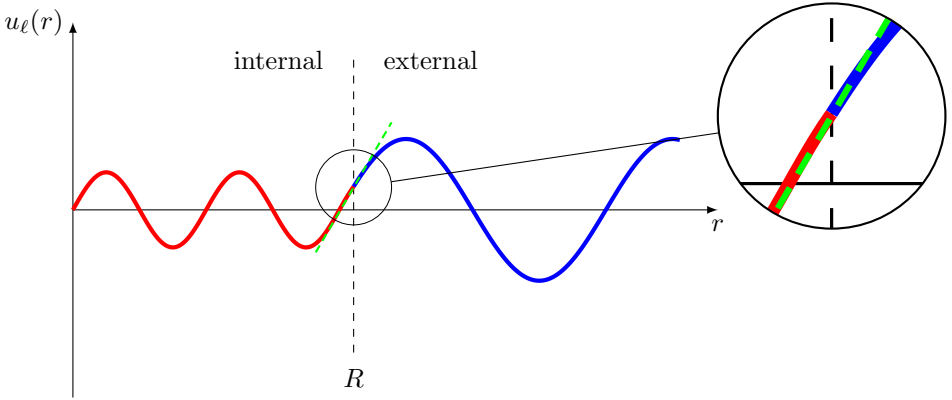
\begin{figure}[!ht]
  \centering
  \pgfmathsetmacro{\w}{205}
\pgfmathsetmacro{\A}{0.5}
\pgfmathsetmacro{\R}{3.7}
\usetikzlibrary{spy}
\begin{tikzpicture}[smooth,samples=100,spy using outlines={circle, magnification=3, size=3cm, connect spies}]
\draw[dashed] (\R,2)--+(0,-4) node[below] {$R$};
\draw (\R,2)  node[left] {internal\phantom{O}} node[right] {\phantom{O}external};
\draw[-latex]  (0,-2.5)--+(0,5) node[left] {$u_{\ell}(r)$};
\draw[-latex] (0,0)--+(8.5,0) node[below] {$r$};
\draw[color=red,ultra thick] plot[domain=0:\R] (\x,{ sin(\w*\x)*\A});
\draw[color=blue,ultra thick] plot[domain=\R:8] (\x,{sin(\w/2*\x)*\A/sin(\w/2*\R)*sin(\w*\R)});

\draw[thick,densely dashed,green] ($(\R,0.3)+(60:-1)$) -- ($(\R,0.3)+(60:1)$);
\spy on (\R,0.3) in node [right]  at (8.5,1.25);

\end{tikzpicture}
  \caption[Schematic representation of the wave function
  radial]{Schematic representation of the radial wave function,
    separating the inner (red) and outer part
    (blue).  \label{fig:wfct:scheme}}
\end{figure}

The value of $\eta_\ell$ is connected to the conditions of continuity
at the surface (see a schematics representation in figure
\ref{fig:wfct:scheme}). To calculate these conditions, we define
$f_\ell$, the logarithmic derivative of the radial part of the wave
function at the surface:
\begin{equation}
  f_\ell = R \left[ \frac{d u_\ell/dr}{u_\ell}\right]_{r=R}  \label{eq:fl}
\end{equation}
For the part outside the core, we can define:
\begin{equation}
  R \left[ \frac{d u_\ell^{(+)}/dr}{u_\ell^{(+)}}\right]_{r=R} =
  \Delta_\ell + i s_\ell \label{eq:fl:out}
\end{equation}
with $\Delta_\ell$ and $ s_\ell$ real, which correspond to the energy
shift and the penetrability:
\begin{IEEEeqnarray}{rCl}
  \Delta_\ell & = & R\ v_\ell\ \left[G_\ell(R)\ G_\ell'(R) + F_\ell(R)\ F_\ell'(R)\right]\\
  s_\ell & = &    R\ v_\ell\ \left[G_\ell(R)\ F_\ell'(R) - F_\ell(R)\ G_\ell'(R)\right] = k R v_\ell
\end{IEEEeqnarray}

$F', G'$ being the derivatives of $F, G$ and:
\begin{equation}
  v_\ell= \frac{1}{G_\ell^2\left( R\right) +F_\ell^2\left( R\right)} = \frac{2/{\pi
     k R}}{
    J_{\ell+1/2}^2(kR) + Y_{\ell+1/2}^2(kR)}
\end{equation}

the penetrability factor itself.  As its name indicates $v_\ell$ is a
measure of how much the neutron penetrates the nucleus. So
$v_\ell \ll 1$ means that the neutron does not penetrate enough,
resulting in a weak interaction.

We  define the phase shift $\xi_\ell$ as:
\begin{equation}
  \exp\left(2 i \xi_\ell\right) =
  \frac{u_\ell^{(+)}(R)}{u_\ell^{(-)}(R)} = \frac{G_\ell(R)-iF_\ell(R)}{G_\ell(R) + i F_\ell(R)}
\label{eq:dephas}
\end{equation}

Finally we can write the link between $\eta_\ell$ and $f_\ell$, which
gives after some rearrangements:
\begin{equation}
  \eta_\ell =
  \frac{f_\ell-\Delta_\ell+is_\ell}{f_\ell-\Delta_\ell-is_\ell}  \exp\left(2 i \xi_\ell\right)
\end{equation}

The equations \eqref{eq:xs:scatterin} and \eqref{eq:xs:reaction} then
provide the scattering and reaction cross-sections:
\begin{IEEEeqnarray}{rCl}
\label{eq:xs:scatt:devel}  \sigma_{sc,\ell} & = & (2\ell+1)\pi \lambdabar^2
  \left\|A^\ell_\mathrm{res}+A^\ell_\mathrm{pot}\right\|^2\\[.5em]
  \text{with } && \label{eq:Ares} A^\ell_\mathrm{res} = \frac{-2\ i\ s_\ell}{\left(\Re(f_\ell)-\Delta_\ell\right)+i\left(\Im(f_\ell)-s_\ell\right)}\\
 \label{eq:Apot} && A^\ell_{pot} = \exp\left(-2 i \xi_\ell\right) -1
\end{IEEEeqnarray}

amplitudes for the internal ``scattering'' (\textit{i.e.}  resonance)
and for the external potential. And:
\begin{IEEEeqnarray}{rCl}
  \sigma_{r,\ell} & = & (2\ell+1)\pi \lambdabar^2\frac{-4\ s_\ell\
    \Im(f_\ell)}{\left(\Re(f_\ell)-\Delta_\ell\right)^{2}+\left(\Im(f_\ell)-s_\ell\right)^{2}} \label{eq:xs:reaction:devel}
\end{IEEEeqnarray}

In these equations, only the expression of $f_\ell$ is missing to
compute the cross-sections, which we will do later with some
approximations. We can nevertheless conclude here, from
\eqref{eq:xs:reaction:devel}, that $f_\ell$ must be zero or
negative. Since $s_\ell$ appears in the numerator, the reaction
cross-section follows the evolution of the penetrability. Finally, by
looking only at the resonant part of the diffusion
($A^\ell_\mathrm{res}$) we find that for given $s_\ell$ and
$\Delta_\ell$, the amplitude is more important as $f_\ell$ is low.

\subsection{Expression of the shape of a resonance}

The main idea, given the results obtained previously, is to find an
expression of $f_\ell$ showing its energy dependence.  We will assume
that we have purely a resonance and that the input channel of the
reaction is equal to the output channel (resonant elastic
scattering). Knowing that $f_\ell$ is the derivative of $u_\ell$ we
have to find an expression of the latter.

Just inside the core ($r \leq R$) we can write, by
definition of the phase shift \eqref{eq:dephas}:
\begin{equation}
  u_\ell(r) \sim \exp(-ikr) + \exp(ikr +2i \xi_\ell) =
  2\ \exp(i\xi_\ell)\ \cos(kr+\xi_\ell)
\end{equation}

We have therefore for its derivative, evaluated in $R$:
\begin{equation}
  u^{\prime}_\ell(R) \sim
- 2\ \exp(i\xi_\ell)\ k\ \sin(kR+\xi_\ell)
\end{equation}

and consequently:
\begin{equation}
 f_\ell(R) = R \frac{ u^{\prime}_\ell(R)}{u_\ell(R)} \sim -kR \tan
 (kR+\xi_\ell) \label{eq:fell:res}
\end{equation}

Recalling that $\xi_\ell$ depends \textit{a priori} on energy, this
function will alternate in energy between two poles and zero
values. This last value maximizes \textit{a priori} the amplitude of
the resonant scattering (equation \eqref{eq:Ares}).

We can therefore expand $f_\ell$ around the resonance energy $E_0$
:
\begin{equation}
 f_\ell(E) = \left(E-E_0\right)\ \underbrace{\left[\frac{d f_\ell(E)}{d E}
 \right]_{E=E_0}}_{f^\prime(E_0)} +\ldots
\end{equation}

The expression for the amplitude \eqref{eq:Ares} becomes:
\begin{equation}
  A^\ell_\mathrm{res} = \frac{-2\ i\ s_\ell}{\left[\left(E-E_0\right) f^\prime(E_0)-\Delta_\ell\right]-i\ s_\ell}
\end{equation}

which can be rewritten:
\begin{IEEEeqnarray}{rCl}
  A^\ell_{res} & = & \frac{i\ \Gamma}{\left(E-E_0 -\Delta E\right)+i\frac{1}{2}\Gamma} \\[1em]
  \text{with} && \gamma^2 = -1/f^\prime(E_0) \label{eq:reduced:width:def1}\\
  & & \Gamma = 2 s_\ell\gamma^2 \label{eq:width:fct:reduced:width}\\
  & & \Delta E = \gamma^2\Delta_\ell
\end{IEEEeqnarray}

The quadratic norm of this amplitude, directly related to the
cross-section, has the well-known form of a dispersion described
by Breit \& Wigner:
\begin{equation}
  \left\|A^\ell_\mathrm{res}\right\|^2 = \frac{\Gamma^2}{\left(E-E^\prime_0
     \right)^2+\frac{1}{4}\Gamma^2}
\end{equation}

where the energy $E^\prime_0 = E_0 -\Delta E$ is the energy of the
resonance shifted by a factor $\Delta E$.

For a concise development of the equations presented here we can refer
the reader for example to F.~Gunsing's lectures at Joliot-Curie School
2014 \cite{Gunsing:EJC2014}.

\subsection{Expression of the energy-dependent apparent width and   position of the resonant state
  \label{sec:th:width:fct:nrj}}


We start from the characteristic expression of the cross-section
distribution as deduced in the previous section:
\begin{equation}
  \frac{d\sigma}{d E} \propto \dfrac {\Gamma_\ell\left(E\right)}{\left( E^\prime_0-E\right)^{2}+\dfrac {1} {4}\Gamma^2_\ell\left( E\right) }
\end{equation}

The expression of the width and the energy shift are valid whatever
$E$ is, so in particular for $E_0$
\begin{equation}\label{eq:widthevo}
  \Gamma_\ell\left( E_0\right) =2s_\ell\left( E_0\right) \gamma^{2}
  = \Gamma_0
\end{equation}

the reduced width $\gamma$ being constructed to be constant in energy we can write:
\begin{equation}
  \frac{\Gamma_\ell\left( E\right)}{\Gamma_\ell\left( E_0\right)}
  = \frac{\Gamma_\ell\left( E\right)}{\Gamma_0}
  = \frac{s_\ell\left( E\right)}{s_\ell\left( E_0\right) }
\end{equation}

and finally by developing the penetrability expression
$s_\ell\left( E\right)$:
\begin{equation}
  \Gamma_\ell\left( E\right) = \Gamma_0 \frac{k}{k_0} \frac{
    J_{\ell+1/2}^2(k_0R) + Y_{\ell+1/2}^2(k_0R)}{
    J_{\ell+1/2}^2(kR) + Y_{\ell+1/2}^2(kR)}
  \label{eq:Gl:Bessel}
\end{equation}

And in the same way for the energy shift:
\begin{equation}
\Delta E = -\left( s_\ell - B\right) \gamma^{2}
\end{equation}

equation \eqref{eq:widthevo} allows us to extract the reduced
width $\gamma^{2} = \Gamma_0/(2 s_\ell\left( E_0\right) )$ which
we reinject:
\begin{equation}
\Delta E = -\Gamma_0\frac{s_\ell\left(E\right) - B}{2 s_\ell\left( E_0\right)}
\end{equation}

$B$ is a constant that we choose such that at the energy of the
resonance the shift is equal to zero, \textit{i.e.}
$B = s_\ell\left(E_0\right)$ and thus:
\begin{equation}
  \Delta E = -\Gamma_0\frac{s_\ell\left(E\right) - s_\ell\left(E_0\right)}{2 s_\ell\left( E_0\right)}
\end{equation}

as for $s_\ell\left(E\right)$ its expression:
\begin{equation}
s_\ell\left(E\right) = R\ v_\ell\ \left[G_\ell(R)\ G_\ell'(R) + F_\ell(R)\ F_\ell'(R)\right]
\end{equation}

can be written, using the different properties of the Bessel functions
:
\begin{align}
   s_\ell\left(E\right) = & \frac{-\ell}{J_{\ell+1/2}^2\left(kR\right)+Y_{\ell+1/2}^2\left(kR\right)} \notag\\
  &
    \left\{\left[J_{\ell+1/2}^2\left(kR\right)\ell+Y_{\ell+1/2}^2\left(kR\right)\right] \right.\notag\\
  & \left.-kR\left[J_{\ell+1/2}\left(kR\right)J_{\ell-1/2}\left(kR\right)+Y_{\ell+1/2}\left(kR\right)Y_{\ell-1/2}\left(kR\right)\right]\right\}
\label{eq:Sl:Bessel}
\end{align}

The forms of the equations \eqref{eq:Sl:Bessel} and
\eqref{eq:Gl:Bessel} are therefore simple enough to be implemented in
a computer code (there are many libraries  for Bessel functions, for
example the GSL \cite{GSL:ref}). In particular their expression is
general whatever  $\ell$ is, allowing to study the dependence in
$\ell$ as an additional parameter, possibly without any prior
hypothesis.

\section{Systematic studies \label{sec:systematic:BW}}

\subsection{General cases}

\begin{figure}[!ht]
  \centering
  \includegraphics[width=0.49\textwidth]{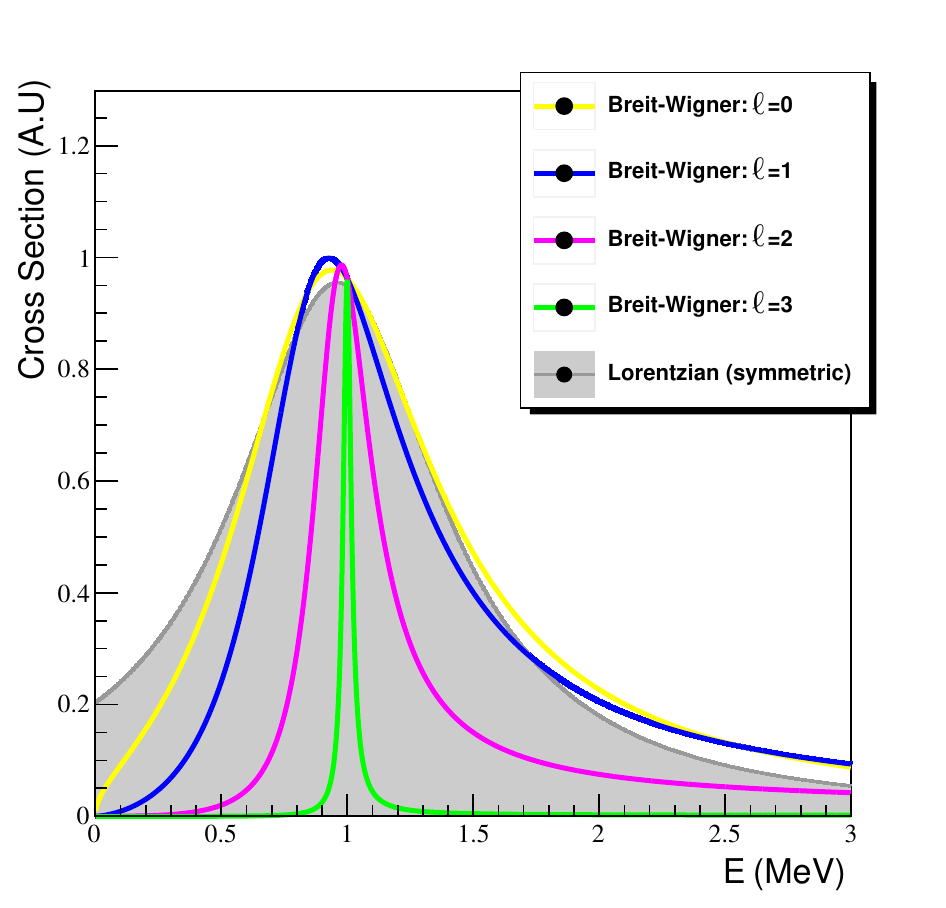}
  \includegraphics[width=0.49\textwidth]{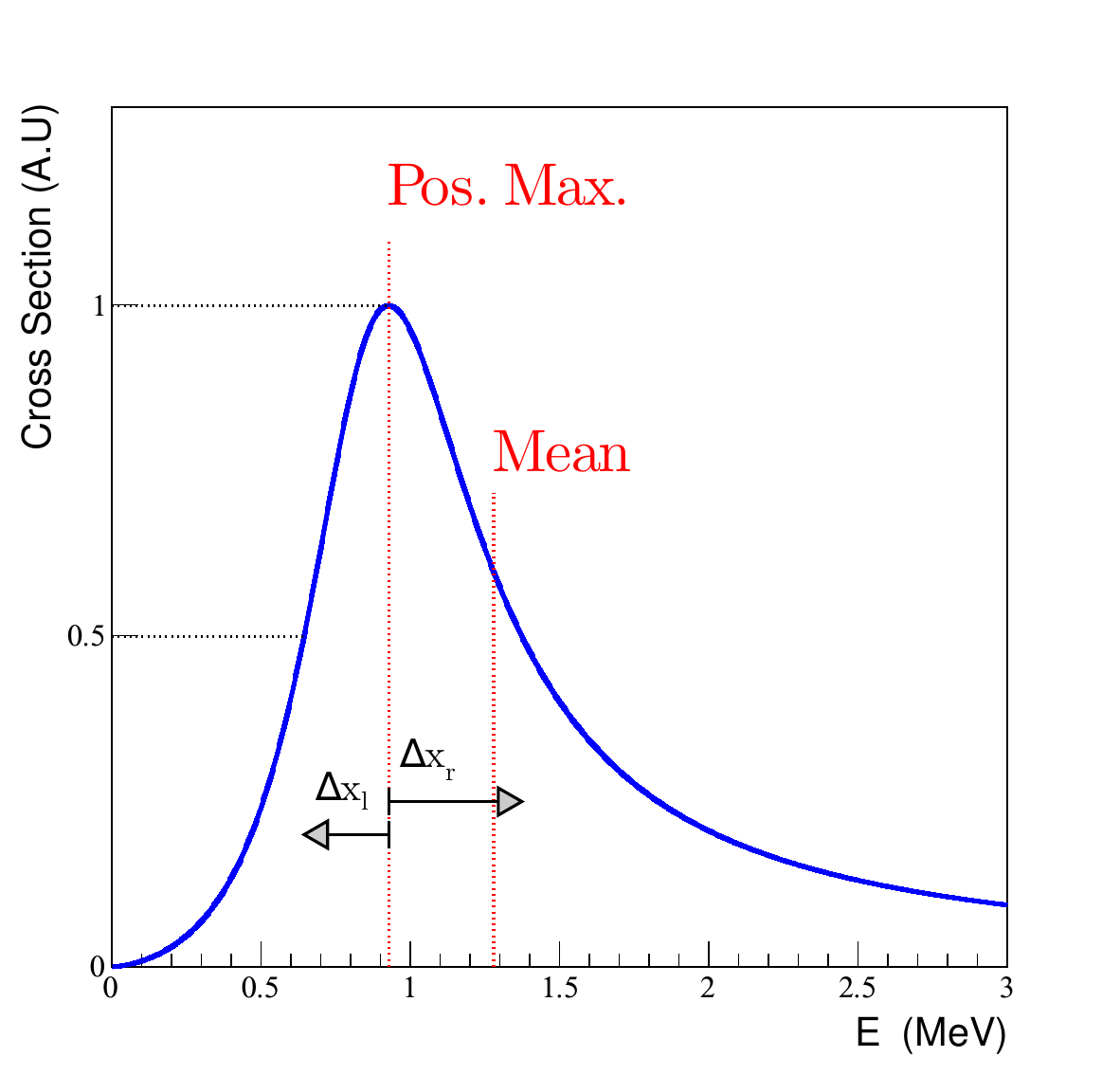}
  \caption [Lorentzian and Breit-Wigner]{Left: 
    Lorentzian (grey) distribution compared to Breit-Wigner
    distributions with 
    energy and $\ell$ dependent widths, taking into account
    the energy shift. For all curves
    $E_0 = \Gamma_0 = \SI{1}{MeV}$, only $\ell=0,1,2$ and $3$
    (respectively yellow, blue, magenta and green) change. We
    assume  here a  $^{17}\text{B}+n$ system. Right: illustration of
    observables used to construct the estimators of figures
    \ref{fig:th:syst:width} and \ref{fig:th:syst:energy}, see text 
    for details.
    \label{fig:th:ex:BW}
  }
\end{figure}

On figure \ref{fig:th:ex:BW} are given the typical shapes of
distributions (in color), compared to a Lorentzian with the same
parameters (in gray), for $E_0 = \Gamma_0 = \SI{1}{MeV}$. We first
recall that these distributions physically represent
cross-sections. It seems quite natural that if the energy $E$ -- which
corresponds in the case of the scattering of a neutron on a fragment
at rest to the kinetic energy of the neutron -- is zero then the
resonance probability is also zero.  At constant integral, if we bring
the cross-section to zero for $E_0=0$ the ``apparent'' maximum of the
function will therefore shift to low energies\footnote{Except for
  $\ell=0$ states, see section \ref{sec:virtual-states}.}.

\begin{figure}[!ht]
  \centering
  \includegraphics[width=0.99\textwidth]{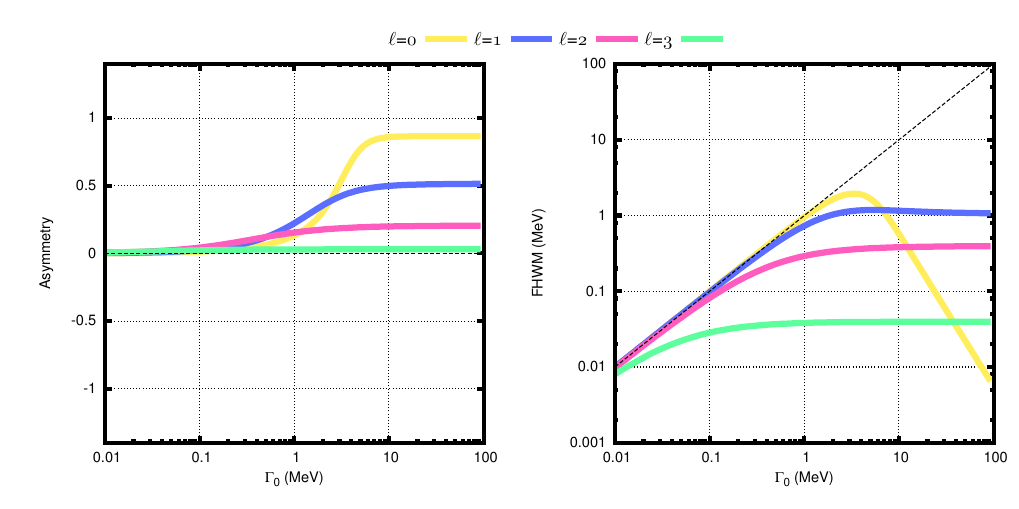}
  \caption [Breit-Wigner characteristics depending on width]{
    Asymmetry (left) and FWHM (right) function of
    the intrinsic width of a Breit-Wigner centered at $E_0 = 1$~MeV for
    $\ell=0,1,2$ and $3$ (respectively yellow, blue, magenta and
    green). We consider here a $^{17}\text{B}+n$ system. The dashed lines correspond to a Lorentzian function.
    \label{fig:th:syst:width}
  }
\end{figure}

\begin{figure}[!ht]
  \centering
  \includegraphics[width=0.99\textwidth]{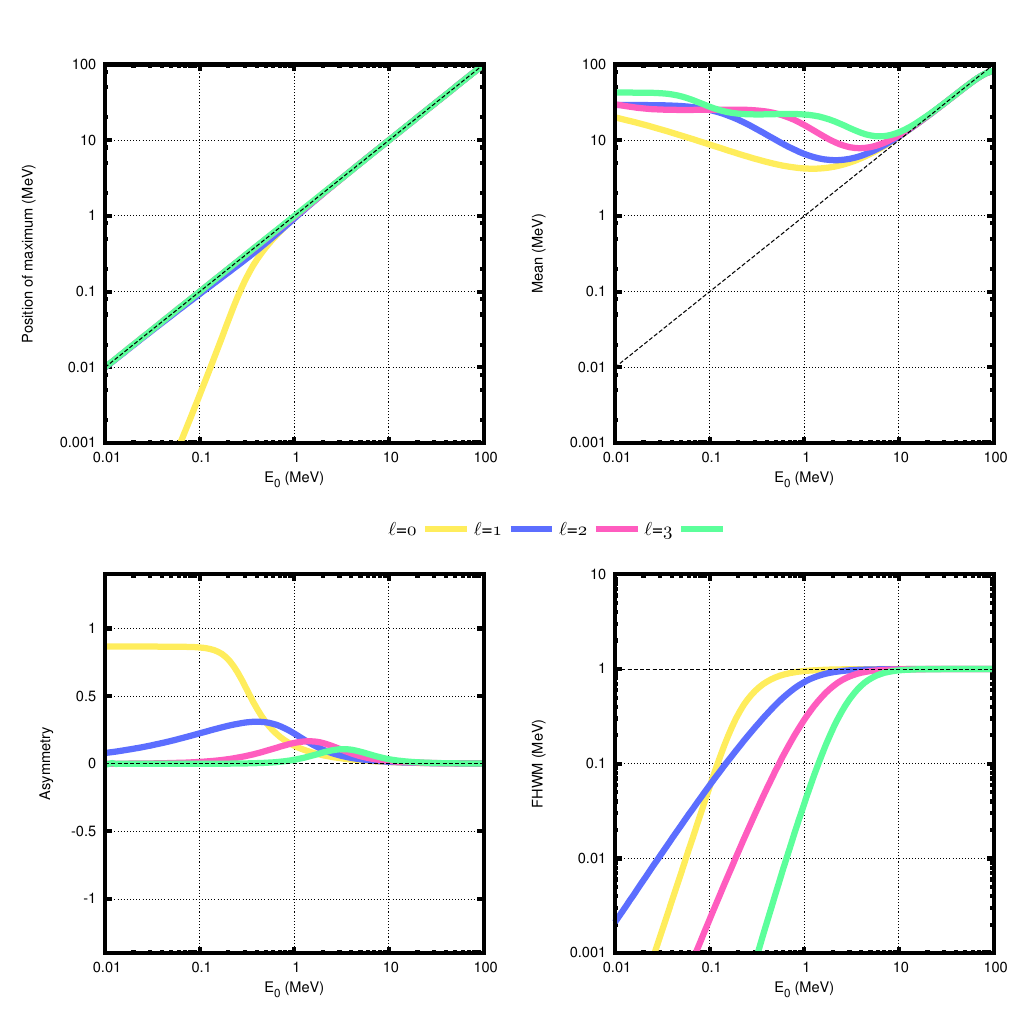}
  \caption[Breit-Wigner characteristics as a function of excitation
  energy]{Position of maximum (top left), mean energy (top right),
    asymmetry (bottom left) and FWHM (bottom right) function of
    energy of a
    Breit-Wigner of width $\Gamma_0 = \SI{1}{MeV}$ for $\ell=0,1,2$ and $3$
    (yellow respectively, blue, magenta and green). We assume a
     $^{17}\text{B}+n$ system.
    \label{fig:th:syst:energy}
  }
\end{figure}

Since taking into account the energy shift and the energy dependence
of the width significantly modifies the distribution compared to a
Lorentzian one, we present on figures \ref{fig:th:syst:width} and
\ref{fig:th:syst:energy} the evolution of these characteristic
distributions as a function of the energies and widths of the
resonances. In order to quantify this evolution, we define several
estimators:
\begin{itemize}
\item the full width half maximum (FWHM), which is directly comparable
  to the $\Gamma$ of the Lorentzian,
\item the position of the maximum,
\item  the mean value,
\item and a measure of the asymmetry.
\end{itemize}

Concerning this asymmetry, we decided not to take the usual Pearson's
coefficient of skewness (or \textit{skewness}\footnote{Mathematically
  the moment of order 3 of the reduced centered variable
  $\gamma_{1}=\operatorname {E} \left[\left(\frac {X-\mu }{\sigma}
    \right)^{3}\right]=\frac{\mu_3}{\sigma^3}$ with
  $\mu_{k}=\operatorname {E} \left[(X-\mu )^k\right]=\int_{-\infty
  }^{+\infty }(x-\mu )^k P(x)\mathrm {d} x$ and $^{2} = \mu_{2}$})
because it was found to be numerically unstable due to the infinite
integration. Given the functions studied (analytic and continuous with
one maximum) we preferred a more graphical version (noted
$\mathcal{A}$) which consists in measuring the ``half''-widths at half
maximum on the right $\Delta x_r$ \emph{and} left of the maximum
$\Delta x_{l}$ (see on figure \ref{fig:th:ex:BW} on the right),
calculate the difference and normalize it to the sum (the \emph{true}
FWHM). Then if we call $\Delta x_\text{r}$ and $\Delta x_\text{l}$ the
half-widths at half height on the right and on the left respectively,
the respectively, so the asymmetry is written
$ \mathcal{A}=\frac{\Delta x_\text{r}-\Delta x_\text{l}}{\Delta
  x_\text{r}+\Delta x_\text{l}}$.

This estimator is therefore zero in case of a symmetrical
distribution, positive if the distribution has a tail towards high
energies and negative in the opposite case.

A simple Lorentzian is symmetric (zero skewness) and its FWHM equals
to $\Gamma_0$. The position of the maximum and the mean are both equal
to $E_0$. All these cases are represented in black dashed lines on
figures \ref{fig:th:syst:width} and \ref{fig:th:syst:energy}.  On our
illustrations the axes are in MeV and we will take as reference a
distribution with $E_0 = \Gamma_0 = \SI{1}{MeV}$. We could also have
put axes ($E/E_0, \Gamma/\Gamma_0$ etc) since these distributions can
be scaled by theses factors. Note that the result is not
\textit{necessarily physical} since certain combinations/values of
$E_0$, $\Gamma_0$ and $\ell$ are not relevant (see section
\ref{sec:sp:width}).

Let's start with the left panel of figure \ref{fig:th:syst:width}
\textit{i.e.}  the asymmetry function of the intrinsic width for a
resonance at $E_0=\SI{1}{MeV}$ for different $\ell$. This one is
always positive (or almost zero for very low $\Gamma_0$) because the
distribution presents a tail always towards high energies. The latter
tends to move away from zero for $\Gamma_0 < E_0$ to saturate at a at
values depending on the $\ell$ (the lower $\ell$ is the lower the
saturation value). Consequently, on the right panel of figure
\ref{fig:th:syst:width} which shows the real width as a function of
the intrinsic width, we observe a saturation of the real width when
the intrinsic width increases. Therefore, \emph{even if experimentally
  our resolution is minimal, we are only marginally sensitive to large
  widths} and this is all the more true as $\ell$ gets larger, up to
``saturation''.  We note that the distributions $\ell=0$ have a
different behavior, in particular the width is maximal around
$\Gamma_0 = E_0$ and decreases for width greater than $E_0$. This is
not surprising given the particular nature of these resonances (see
section \ref{sec:virtual-states}).

Concerning the evolution with the intrinsic energy of the resonance
(at constant intrinsic width $\Gamma_0=\SI{1}{MeV}$), the different
estimators are presented on figure \ref{fig:th:syst:energy}.  We first
point out that these are not necessarily independent of each other.
We note in the upper right corner that except for $\ell=0$ the
position of the maximum almost systematically coincides with
$E_0$. The asymmetry (bottom left) is significant at low $E_0$ then
tends to $0$ at large $E_0$, which is the counterpart of the left
panel of figure \ref{fig:th:syst:width}: it is equivalent to say that
the larger the intrinsic width is compared to the energy of the
resonance, the greater the asymmetry. This last observation has a
direct consequence on the mean (top right), which is strongly shifted
from the maximum to $E_0 \ll \SI{1}{MeV}$. As for the FWHM, it is also
away from \SI{1}{MeV} in the same range of conditions.

\subsection{Virtual states\label{sec:virtual-states}}

The previous systematic studies show that $\ell=0$ states follow
different trends with respect to $\ell>0$ states. This behavior leads
to what is called ``virtual states'' and is well explained by McVoy in
his article \cite{McVoy1968}. To understand it, let's look at the
evolution with energy of the widths of the states, as given by the
equation \eqref{eq:width:fct:reduced:width}. The latter shows the
``penetrability'' $s_\ell$ whose expression is recalled here:
\begin{equation}
  s_\ell=     \frac{2/\pi}{J_{\ell+1/2}^2(kR) + Y_{\ell+1/2}^2(kR)}
\end{equation}
The Bessel functions $J_p(x)$ are written as:
\begin{equation}
  J_p\left(x\right) =\sum^{\infty }_{m=0}\frac {\left( -1\right)
   ^{m}}{m!\ \mathbf{\Gamma}\left(m+p +1\right)}\left(\frac{x}{2}\right)^{2m+p}
\end{equation}
with $\mathbf{\Gamma}$ the gamma function. The Bessel functions
$J_p$ and $Y_p$ being related, for $p = \ell + 1/2$ with $\ell$ integer
positive, by:
\begin{equation}
  J_{-(\ell+1/2)}(x) = (-1)^{\ell+1}    Y_{\ell+1/2}(x)
\end{equation}

this implies that $s_\ell$, and thus by extension the width $\Gamma$,
evolves in $k^{2\ell+1}$. This is to be compared to the energy which
varies in $k^{2}$ (equation \eqref{eq:def:k}) and so for $\ell>0$ the
width decreases faster than the energy. In this case the state never
overlaps the threshold, so there can not be any ambiguity as to
whether it occurs above the threshold (a resonance) or below (a bound
state).  If $\ell =0$ however, the width decreases \emph{only} as the
square root of the energy, so as the energy decreases, there will
necessarily be an ``overlap'' threshold before it becomes a bound
state. In some cases it is not clear whether the level is
predominantly above or below the threshold, and to describe this
situation the term ``virtual'' state is used.  In order to illustrate
this situation, we present the evolution of the width as a function of
the energy, taking as reference the energy of the resonance
$E_0 = \SI{1}{MeV}$ (from the equation \eqref{eq:Gl:Bessel} and with the
same logic as the previous systematic analyses). We can see that only
for $\ell = 0$ (in yellow) the width value is greater than the energy
(in dotted line).
\begin{figure}[!ht]
  \centering
  \includegraphics[width=0.49\textwidth]{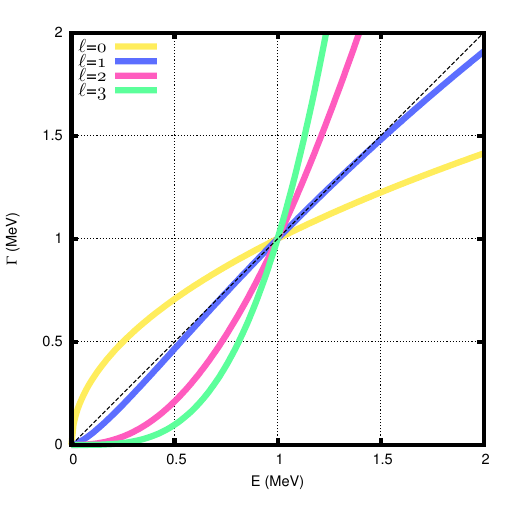}
  \caption[Evolution of the width with energy]{Evolution of
    the width as a function of the energy of the resonance, for
    $\ell=0,1,2$ and $3$ (respectively yellow, blue, magenta and
    green). We consider  here a $^{17}\text{B}+n$ system  with
    $E_0 = \SI{1}{MeV}$. The $\Gamma = E$ case is depicted as
    a dotted black line.
    \label{fig:BW:widths:w:energy:virtual}
  }
\end{figure}

\subsection{Width  of single-particle resonances
  \label{sec:sp:width}}

Experimentally the width of an independent particle state is often
used, knowing the nature of the considered state, to deduce its
spectroscopic factor. The link between the two can simply be seen in
the definition, here simplified, of the reduced width as given in
references \cite{rmatrix,Lane1960}:
\begin{IEEEeqnarray}{rrrCl}
  & & \gamma_{\ell,i} & = & \frac{\hbar}{\sqrt{2\mu R}}\int \phi_\ell^{\star}
  \chi_i dS \label{eq:reduced:width:sp} \\
  \text{where } &:&\ dS && \text{surface element at } r=R\\\nonumber 
  & & \phi && \text{surface component of the internal wave function}\\\nonumber
  & & \chi_i && \text{internal eigenfunction, for an given energy
 $E_{i}$}\\\nonumber
\end{IEEEeqnarray}

The surface component corresponds to what must be multiplied with the
radial wave function $u_\ell$ to obtain the total wave function
$\psi$:
\begin{equation}
  \label{eq:surface:function}
  \psi = \sum_\ell u_\ell \ \phi_\ell = \sum_i A_i \chi_i
\end{equation}

Quantitatively, it appears here because $\gamma_\ell$, if we refer to
equation \eqref{eq:reduced:width:def1}, shows the derivative according
to $r$ of the wave function. Note that in expression
(\ref{eq:surface:function}) presented here the summation is simplified
on $\ell$ but, in particular $R$--matrix formalism
\cite{rmatrix,Lane1960}, one must take into account all the quantum
numbers of the input channels of the reaction ($j,l,J$, etc).
 
We clearly see in this expression that the spectroscopic factor
$C^{2}S$, is directly connected to the observed width. Schematically:
\begin{equation}
  \label{eq:facteur:spectro}
  C^{2}S = \sigma_{\ell,\text{exp}} / \sigma_{\ell,\text{theo}}
\end{equation}

with $\sigma_\ell$ the cross-section of interest,
$\sigma \propto \left\|\phi\right\|^{2}$, experimental and theoretical
respectively. So schematically again we can write
$\phi_\text{exp} = \sqrt{C^{2}S}\ \phi_\text{theo}$, which gives:
\begin{equation}
  \label{eq:facteur:spectro:width}
  C^{2}S = \gamma_{\ell,\text{exp}} / \gamma_{\ell,\text{theo}}
\end{equation}

Now to really evaluate the width of interest it is necessary to
calculate the corresponding wave functions, for example from
shell-model calculations, which is beyond this simple
introduction. However, with a minimum of assumptions, notably a square
well, it is possible to estimate these widths. Bohr and Mottelson then
derive \cite[page~440]{bohr:mott}: 
\begin{equation}
\Gamma_{sp} =
\begin{cases}
\frac{2\ \hbar^{2}}{\mu R^{2}}\ k R\ v_\ell\ \frac{2\ell-1}{2\ell+1} &
\text{if } \ell>0 \text{ and }   kR < \ell^{1/2},\\
\frac{2\ \hbar^{2}}{\mu R^{2}}\ k R & \text{if } \ell = 0
\end{cases}
\label{eq:gamma:sp:BM}
\end{equation}
\begin{figure}[!ht]
  \centering
  \includegraphics[width=0.49\textwidth]{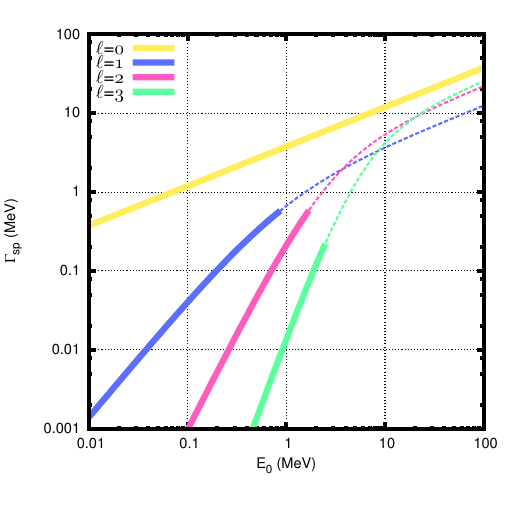}
  \caption[Independent particle width, function of its energy]{Independent particle width as a function of energy
     of the resonance, according to the formalism of Bohr \& Mottelson
    \cite{bohr:mott} and for $\ell=0,1,2$ and $3$ (respectively yellow, blue,
    magenta and green). We assume a  $^{17}\text{B}+n$ system. The
    dotted lines correspond to the case outside the necessary
    conditions for $kR < \ell^{1/2}$.
    \label{fig:BW:widths:dependence}
  }
\end{figure}

A systematic study, using this formalism, is presented in
figure~\ref{fig:BW:widths:dependence} for the system $^{17}\text{B}+n$,
with the same conventions as in the section \ref{sec:systematic:BW},
despite the limits of definition of \eqref{eq:gamma:sp:BM}, which
explains why some values are ``missing'' (but presented here for
information as a dotted line).  We see in particular that the larger
$\ell$ the smaller the width, which comes from the penetration factor
effect.

\subsection{Multineutron resonances}

\begin{figure}[!ht]
  \centering
  \includegraphics[width=0.49\textwidth]{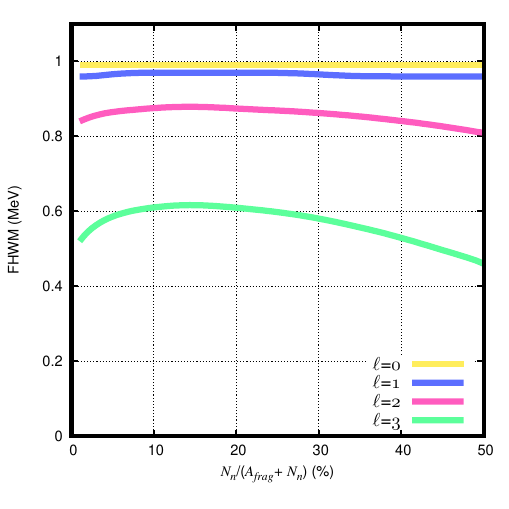}
  \includegraphics[width=0.49\textwidth]{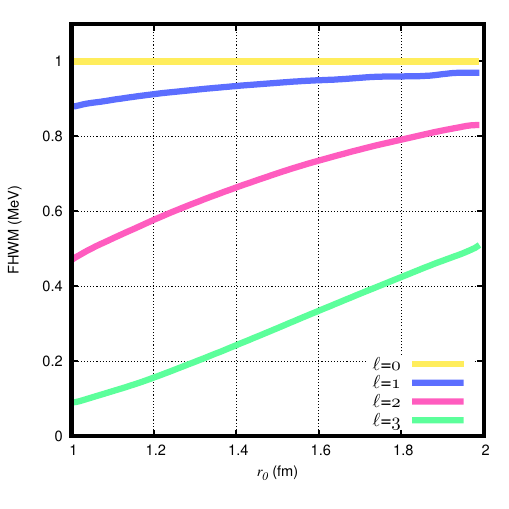}
  \caption {Left: dependence of the width for two-body resonances, for
    different $\ell$, function of the fraction of incident clustered
    neutrons.  Right: same plot but function of the $r_0$ parameter
    used to calculate the fragment radius.  \label{fig:width:xn:ro}}
\end{figure}

The previous section presented two-body resonances and mainly
fragment--neutron ones. Simple parametrizations for multi-body cases
and three-body in particular are difficult because one of the
assumption of the traditional $R$--matrix approach is the ``absence or
unimportance of all processes in which more than two product nuclei
are formed'' \cite{rmatrix} which means that the theory cannot be used
immediately for at least three-body decays. More complex modelization
and theoretical calculations are thus necessary. One of the
difficulties, both theoretical and experimental, is that several decay
paths and combinations of them open. We should note however that some
attempts have been done in particular when the decay is sequential
\cite{Barker2003,Fynbo2009}, which in this case can be regarded as a
two-step two-body process.

We would like to draw the attention on the fact that in absence of
specific calculations authors generally extract the energy and the
width of the ``$N$-body'' resonance using the parametrization
described in the previous sections for two bodies. They sometimes also
assume that $N_n$ clustered neutrons participate to the resonance, as
a multineutron ``bound'' system.  In the authors' opinion, in absence
of any other model, and as far as the resonance shape does not
resemble a virtual one, the only acceptable parametrization to be used
is $\ell=0$ because this one does not exhibit any energy dependence on
the position and the width. As shown also on figure
\ref{fig:width:xn:ro}~(left) there is also no dependence in the width
with the $N_n$ neutrons clustered in a two-body resonance. Most of
these systems close to the drip line present a substantial extension
(halo, skin\ldots) and a study of the effect of the radius $r_0$
(figure \ref{fig:width:xn:ro}, right) also shows that $\ell=0$
distributions are independent of this parameter. One would have to be
careful with the exact meaning of this width which, even if the
experimental resolution is taken into account to extract it, would be
difficult to compare to theory. Probably the best approach will then
be to produce from the theoritical approach the energy distribution
and then fit it with the same $\ell=0$ distribution.

\section{Conclusion}

We presented in this paper a series of simple calculations to describe
two-body resonances and alike. These results represent a
parametrization often used to extract the properties of resonances
measured experimentally, in particular for nuclei far from stability
or beyond drip-lines. The main purpose of this article was to put then
in perspective and illustrate the evolution of the measurable
properties with their intrinsic parameters, in particular to the
attention of physicists studying resonance properties and comparing
them to theoretical models.



\bibliography{gibelin.bib}
\end{document}